%% file: root.tex
\documentclass[letterpaper, 10 pt, conference]{ieeeconf}  

\IEEEoverridecommandlockouts                              
\usepackage{amsmath} 
\usepackage{amssymb}  
\usepackage{mathtools}
\usepackage{url}
\usepackage{bbm}
\usepackage{courier}
\usepackage{algorithmicx}
\usepackage{subcaption}  
\usepackage{cleveref}
\usepackage{algorithm}
\usepackage[noend]{algpseudocode}

\usepackage{xcolor}
\usepackage{siunitx}
\usepackage{booktabs}

\def\baselinestretch{0.97}
\AtBeginDocument{
}

\input{commands}
\newcommand{\rev}[1]{{\color{black} #1}}

\newcommand{\revf}[1]{{\color{black} #1}}

\title{\LARGE \bf
Distributionally Robust Control with Constraints\\
on Linear Unidimensional Projections 
}

\author{Alexandros E. Tzikas,$^{1}$ Lukas Fiechtner,$^{2}$ Arec Jamgochian,$^{1,3}$ and Mykel J. Kochenderfer$^{1}$
\thanks{$^{1}$A. E. Tzikas (corresponding author), A. Jamgochian, and M. J. Kochenderfer are with the Department of Aeronautics and Astronautics, Stanford University, Stanford, CA 94305, U.S.A.
        {\tt\small \{alextzik, arec, mykel\}@stanford.edu}}
\thanks{$^{2}$L. Fiechtner is with the Institute of Computational and Mathematical Engineering, Stanford University, Stanford, CA 94305, U.S.A.
        {\tt\small fiechtlb@stanford.edu}}
\thanks{$^{3}$TerraAI, Redwood City, CA 94063, U.S.A.}
\thanks{Toyota Research Institute (TRI) provided funds to assist the authors with their research, but this article solely reflects the opinions and conclusions of its authors and not TRI or any other Toyota entity. The NASA University Leadership Initiative (grant $\#$80NSSC20M0163) provided funds to assist the first author with their research, but this article solely reflects the opinions and conclusions of its authors and not any NASA entity. For the first author, this work was also partially funded through the Alexander S. Onassis Foundation Scholarship program.
}
}

\begin{document}

\maketitle
\thispagestyle{empty}
\pagestyle{empty}

\begin{abstract}
Distributionally robust control is a well-studied framework for optimal decision making under uncertainty, with the objective of minimizing an expected cost function over control actions, assuming the most adverse probability distribution from an ambiguity set. We consider an interpretable and expressive class of ambiguity sets defined by constraints on the expected value of functions of one-dimensional \rev{linear} projections of the uncertain parameters. 
\rev{Prior work has shown that, under conditions, problems in this class can be reformulated as finite convex problems. In this work, we propose two iterative methods that can be used to approximately solve problems of this class in the general case.} The first is an approximate algorithm based on best-response dynamics. The second is an approximate method that first reformulates the problem as a semi-infinite program and then solves a relaxation. 
\rev{We apply our methods to portfolio construction and trajectory planning scenarios.}

\end{abstract}

\input{sections/intro}
\input{sections/problem_statement}

\input{sections/baseline}

\input{sections/approach}

\input{sections/results}
\input{sections/conclusion}

\bibliographystyle{IEEEtran}
\bibliography{sislstrings,sample}  

\end{document}

%% file: commands.tex
\newcommand{\st}{\text{ s.t. }}

%% file: sections/intro.tex
\section{Introduction}

In this work, we consider distributionally robust control (DRC), where the unknown parameters \rev{of the system} follow an unknown probability distribution. Although the distribution is not known, we assume that it belongs in a known set \rev{of probability distributions}, called the ambiguity set. 
\rev{This formulation is} suited for applications where the probability distribution \rev{of the parameters} cannot be easily determined or can change over time. \rev{Examples include constructing a portfolio without knowledge of the future asset price movements or fixing the energy production of a power plant before the demands are realized \cite{zhen2023unified}.}

\rev{
In contrast to DRC, classical control under uncertainty often assumes a known probability distribution for the uncertain parameters of the system. For example, in stochastic optimization, empirical risk minimization (ERM) optimizes the expectation of an objective function $l$, which estimates costs given parameters $x$, distributed according to a known distribution $\mathbb{P}$, and design inputs $u$ in set $\mathcal{U}$, \rev{with} $u^*_{\text{ERM}} = \arg\min_{u\in\mathcal{U}} \mathbb{E}_{x \sim \mathbb{P}} [l(u,x)]$~\cite{boyd2009convex}. In practice, assuming a known probability distribution for the parameters can be fragile\rev{;} if the true distribution deviates from the assumed model, the solution's performance can deteriorate severely.
Robust control addresses this by optimizing against the worst-case parameter realization within a prespecified set $\mathcal{X}$ \rev{with} $u^*_{\text{robust}} = \arg\min_{u\in\mathcal{U}}\max_{x \in \mathcal{X}} l(u, x)$~\cite{ben2002robust,bertsimas2011theory}. Unfortunately, specifying a complete and accurate parameter set $\mathcal{X}$ a priori can be impractical. Additionally, worst-case optimization can be overly pessimistic against the real parameters.
}

\rev{DRC} bridges the gap between stochastic and robust \rev{control} by considering uncertainty in the probability distribution of the parameters \rev{\cite{zhen2023unified}}. Rather than assuming a known distribution or set for the parameters, DRC instead assumes that the parameters follow a distribution that is contained in an ambiguity set \rev{$\mathcal{A}$} of distributions. This distributional uncertainty set is often chosen to be a neighborhood around a nominal (e.g. empirical) distribution $\hat{\mathbb{P}}$. This yields a minimax problem \rev{where the decision maker} chooses a decision \rev{input} \rev{that minimizes the expected} cost, assuming \rev{a worst-case distribution $\mathbb{P}$}:
\begin{equation}
    u^*_{\text{DRO}} = \arg\min_{u\in\mathcal{U}} \max_{\mathbb{P} \in \mathcal{A}}\mathbb{E}_{x \sim \mathbb{P}}[l(u, x)]\text{.} 
\end{equation}
By optimizing against the worst-case distribution, DRC provides distributional robustness without being overly pessimistic or prescriptive about uncertainty~\cite{rahimian2022frameworks, lin2022distributionally}.

Solution methods for DRC typically reformulate the problem as a semi-infinite program
satisfying constraints over all members of $\mathcal{A}$. They then dualize the problem~\cite{delage2010distributionally,wiesemann2014distributionally}, or iteratively relax the constraints into representative atomic subsets of $\mathcal{A}$~\cite{pflug2007ambiguity,rahimian2019identifying}. 
Typical choices for the ambiguity set consider distributions within an $\epsilon$-ball around a nominal distribution $\hat{\mathbb{P}}$, \rev{when} $\mathcal{A}$ \rev{is defined as} $ \{\mathbb{P} : D(\mathbb{P},\hat{\mathbb{P}}) \leq \epsilon\}$, for some discrepancy measure $D$, such as KL divergence~\cite{ben2013robust,hu2013kullback}, other general $f$-divergences~\cite{namkoong2016stochastic,duchi2021learning}, or Wasserstein distances~\cite{kuhn2019wasserstein, gao2024wasserstein}. 
Efficient solutions to DRC have been considered for many ambiguity sets~\cite{rahimian2022frameworks,lin2022distributionally}. 
A common challenge \rev{in the choice of the} ambiguity set that remains, however, is the selection and interoperability of hyperparameters (e.g. the discrepancy radius $\epsilon$), which can significantly affect both robustness and conservatism.

\textit{In this work, we identify a specific class of problems in DRC, where the constraints on the probability distribution of the parameters involve expectations of functions of one-dimensional \rev{linear} projections of the unknown parameters. We show that this problem class is expressive and the constraint set intuitive enough to be determined in practical applications. \rev{This simplifies the process of selecting the hyperparameters for the ambiguity set.}} \rev{Prior work has shown that if strong duality holds for a problem in this class, then it can be reformulated as a finite convex problem, which can be easily solved under mild conditions (e.g., when the conjugate function can be easily evaluated) \cite{zhen2023unified}.} \textit{We propose two iterative solution methods to approximately solve the problems of this class in the general case. It turns out that, under conditions, these methods involve problems that can be solved using convex optimization \cite{boyd2009convex} or related heuristics \cite{shen2016disciplined}.} 


\rev{This paper is organized as follows. In \cref{sec:statement}, we introduce the problem class of interest and highlight a few important problems within. We then introduce our two proposed solution methods. In \cref{sec:baseline}, we propose a simple approach to solve problems in this class using best-response dynamics. In \cref{sec:approach}, we propose a second solution method that amounts to iteratively solving a semi-infinite program and is based on duality theory and cutting-set algorithms. Next, we present our results and the paper concludes with a brief discussion.}

%% file: sections/problem_statement.tex
\section{Problem Statement}\label{sec:statement}
The problem class we consider consists of optimization problems of the form
\begin{equation}\label{eq:main}
\begin{aligned}
    \min_{u \in \mathcal{U}}\ \max_{\mathbb{P} \in \mathcal{S}}\ &\mathbb{E}_{x \sim \mathbb{P}}\left[ l(u, x) \right]\\
    \mathrm{s.t.}\ &\mathbb{E}_{x \sim \mathbb{P}} \left[ g_i(q_i^\top x) \right] \leq 0\ \mathrm{for\ all}\ i,
\end{aligned}
\end{equation}
\rev{where the input $u$ is constrained \rev{to} $\mathcal{U}$ and the distribution of the random variable $x$, $\mathbb{P}$, must belong in $\mathcal{S}$, where $\mathcal{S}$ is the set of probability distributions with support $\mathcal{X}$.}
The probability distribution $\mathbb{P}$ is \rev{further} constrained by a set of inequalities. Each inequality constrains the expectation of a function of a one-dimensional \rev{linear} projection of $x$. 
We call this class of problems \textit{DRC-1DP} (DRC with 1-Dimensional Projections). 

\rev{We next show that the class DRC-1DP admits useful optimization objectives and interpretable and practical constraints. Consider the problem}
\begin{equation}\label{eq:prototype}
\begin{aligned}
    \min_{u \in \mathcal{U}}\ \max_{\mathbb{P} \in \mathcal{S}}\ &\mathbb{E}_{x \sim \mathbb{P}}\left[ l(u, x) \right]\\
    \mathrm{s.t.}\ &|\mathbb{E}_{x \sim \mathbb{P}} \left[ g_i(q_i^\top x) \right] -p_i| \leq \epsilon_i\ \mathrm{for\ all}\ i
\end{aligned}
\end{equation}
\rev{that belongs in DRC-1DP.}

Some common choices for the objective $l$ include bilinear or convex-concave functions. In the bilinear case, $l(u, x)$ is linear in both $u$ and $x$. This objective has received much attention in the game theory literature \cite{li2022convergence}. In the convex-concave case, $l(u,x)$ is convex in $u$ and concave in $x$ and the minmax theorem \cite{goktas2021convex} holds: $\min_u \max_x l(u,x) = \max_x \min_u l(u,x)$.
 
The ambiguity set of problem \eqref{eq:prototype} consists of all probability distributions with support $\mathcal{X}$ that have an expected value of $g_i(q_i^\top x)$ centered around $p_i$, for all $i$. To demonstrate the interpretability of the ambiguity set of problem \eqref{eq:prototype}, we consider specific cases for $g_i$ below. 
\begin{itemize}
    \item If $g_i(z) = \mathbbm{1}\lbrace z \geq b_i\rbrace $, then $\mathbb{E}_{x \sim \mathbb{P}} \left[ g_i(q_i^\top x) \right] = \mathbb{P} \lbrace q_i^\top x -b_i \geq 0 \rbrace$, and we are constraining the probability content on a selected half-space defined by $(q_i, b_i)$.
    \item If $g_i(z) = z$, then $\mathbb{E}_{x \sim \mathbb{P}} \left[ g_i(q_i^\top x) \right] = \mathbb{E}_{x \sim \mathbb{P}} \left[ q_i^\top x \right]$, and we are constraining the expected value of the one-dimensional projection $q_i^\top x$.
    \item If $g_i(z) = z^p$, then $\mathbb{E}_{x \sim \mathbb{P}} \left[ g_i(q_i^\top x) \right] = \mathbb{E}_{x \sim \mathbb{P}} \left[ (q_i^\top x)^p \right]$, and we are constraining the $p$-th central moment of the one-dimensional random variable projection $q_i^\top x$. 
    \item By combining constraints for $g_i(z) = z^2$ and $g_j(z)=z$ with $q_i=q_j$, we are effectively constraining the variance of $q_i^\top x$.
\end{itemize}
Constraints of the form $|\mathbb{E}_{x \sim \mathbb{P}} \left[ g_i(q_i^\top x) \right] -p_i| \leq \epsilon_i$ for functions $g_i$ in the list above characterize properties of the random vector $x$ that are easily observable and interpretable in practice. For example, if we know that the coordinate $j$ of vector $x$, $x_j$, is positive, we can let $g_i(z)=\mathbbm{1}\lbrace z \geq 0\rbrace$, $q_i$ be the unit vector in the $j$th direction, $p_i=1$, and $\epsilon_i$ be very small.  \rev{The presence of the absolute value in the constraints of \eqref{eq:prototype} and the ability to include higher moment constraints means that \eqref{eq:prototype} does not belong in the class of problems considered in prior work \cite{delage2010distributionally}.}

%% file: sections/baseline.tex
\section{Best Response Algorithm}\label{sec:baseline}
An approximation of the main problem \eqref{eq:main} is
\begin{equation}\label{eq:main_sample}
\begin{aligned}
    \min_{u \in \mathcal{U}} \max_{x^{(1)}, \dots, x^{(n)} \in \mathcal{X}}\ &\dfrac{1}{n}
    \sum_{k=1}^n l(u, x^{(k)}) \\
    \mathrm{s.t.}\qquad &\dfrac{1}{n}\sum_{k=1}^n g_i(q_i^\top x^{(k)}) \leq 0\ \mathrm{for\ all}\ i,
\end{aligned}
\end{equation}
\rev{where the set of points $\lbrace x^{(1)}, \dots, x^{(n)}\rbrace$ defines a discrete probability distribution $\mathbb{P}$, with each point assumed equiprobable. In this case, the expectations simplify to sums of $n$ terms, where each term corresponds to a point in $\lbrace x^{(1)}, \dots, x^{(n)}\rbrace$. Solving problem \eqref{eq:main_sample} amounts to choosing the best input under the worst placement of $n$ points in $\mathcal{X}$. }

We propose to solve problem \eqref{eq:main_sample} using an iterative scheme, as described in \cref{alg:baseline}. 
The algorithm iterates over minimizing the cost under a \rev{fixed discrete distribution} for $x$ and then refining the \rev{discrete distribution of equiprobable points} in a way that maximizes the expected cost. In other words, we choose \rev{a discrete distribution for} $x$ in order to maximize cost and we choose the input $u$ in order to minimize the cost. This is an instance of the best-response algorithm, which is common in game-theoretic settings \cite{kochenderfer2022algorithms}. 

The performance of \cref{alg:baseline} depends on the number of samples $n$. This needs to be sufficiently large to allow for expressivity and feasibility of the constraints in \eqref{eq:sampling_prob2}.
\rev{Finally, we note that there are instances where best-response dynamics do not converge \cite{roughgarden2016twenty}. }

\begin{algorithm}
\caption{\rev{Best-Response} Method}\label{alg:baseline}
\begin{algorithmic}[1]
\Require discrete set $\hat{\mathcal{X}}=\lbrace \hat{x}^{(1)}, \dots, \hat{x}^{(n)}\rbrace \subseteq \mathcal{X}$ of samples

\Repeat
\State Solve 
\begin{equation}\label{eq:sampling_prob1}
    \min_{u \in \mathcal{U}} \dfrac{1}{n}\sum_{k=1}^n l(u, \hat{x}^{(k)})
\end{equation}
and obtain the optimal control $\hat{u}^*$

\State Solve
\begin{equation}\label{eq:sampling_prob2}
\begin{aligned}
    \max_{x^{(1)}, \dots, x^{(n)} \in \mathcal{X}} &\dfrac{1}{n}\sum_{k=1}^n l(\hat{u}^*, x^{(k)})\\
    \st\quad &\dfrac{1}{n}\sum_{k=1}^n g_i(q_i^\top x^{(k)}) \leq 0\ \mathrm{for\ all}\ i
\end{aligned}
\end{equation}
\rev{and assign the optimal variable values $x^{(1), *}, \dots, x^{(n), *}$ to be the points $\hat{x}^{(1)}, \dots, \hat{x}^{(n)}$.}

\Until{convergence}

\State \Return the optimal control $\hat{u}^*$

\end{algorithmic}
\end{algorithm}

%% file: sections/approach.tex
\section{Approach using Semi-Infinite Programming}\label{sec:approach}
We show how to reformulate problem \eqref{eq:main} into a semi-infinite program (SIP) and propose an approximate algorithm to solve the latter.

\subsection{Semi-Infinite Programming Formulation}\label{sec:approach-sip}
Suppose $u$ is fixed. Then, the inner optimization problem of problem \eqref{eq:main}
\begin{equation}\label{eq:inner}
\begin{aligned}
    \max_{\mathbb{P} \in \mathcal{S}}\ &\mathbb{E}_{x \sim \mathbb{P}}\left[ l(u, x) \right]\\
    \st &\mathbb{E}_{x \sim \mathbb{P}} \left[ g_i(q_i^\top x) \right] \leq 0\ \mathrm{for\ all}\ i
\end{aligned}
\end{equation}
can be equivalently written as 
\begin{equation} \label{eq:innerprimal}
    \max_{\mathbb{P}\in \mathcal{S}} \min_{\lambda \geq 0}\ \mathbb{E}_{x \sim \mathbb{P}}\left[ l(u, x) \right] - \sum_i \lambda_i \mathbb{E}_{x \sim \mathbb{P}} \left[ g_i(q_i^\top x) \right].
\end{equation}

Assuming strong duality holds, then problem \eqref{eq:innerprimal} is equivalent to
\begin{equation} \label{eq:innerdual}
    \min_{\lambda \geq 0}\ \max_{\mathbb{P}\in \mathcal{S}} \mathbb{E}_{x \sim \mathbb{P}}\left[ l(u, x) - \sum_i \lambda_i g_i(q_i^\top x) \right].
\end{equation}
Under technical conditions, strong duality holds between \eqref{eq:innerprimal} and \eqref{eq:innerdual} \cite[Theorem 11]{zhen2023unified}. 

We can write problem \eqref{eq:innerdual} in epigraph form
\begin{equation} \label{eq:innerdualepi}
\begin{aligned}
    &\min_{\lambda \geq 0, t}\qquad t\\
    &\st \max_{\mathbb{P}\in \mathcal{S}}\ \mathbb{E}_{x \sim \mathbb{P}}\left[ l(u, x) - \sum_i \lambda_i g_i(q_i^\top x) \right] \leq t
\end{aligned}
\end{equation}
and observe that 
\begin{equation}
\begin{aligned}
    &\max_{\mathbb{P}\in \mathcal{S}}\ \mathbb{E}_{x \sim \mathbb{P}}\left[ l(u, x) - \sum_i \lambda_i g_i(q_i^\top x) \right] - t\leq 0 \Leftrightarrow \\
    &\max_{x \in \mathcal{X}}\ l(u, x) - \sum_i \lambda_i g_i(q_i^\top x) - t\leq 0.
\end{aligned}
\end{equation}
The left implication is obvious. The right implication can be shown by contradiction. Suppose the top condition holds and there exists an $x \in \mathcal{X}$ such that $l(u, x) - \sum_i \lambda_i g_i(q_i^\top x) - t > 0$. Then there exists a probability distribution with support $\mathcal{X}$ that only associates probability mass with the values of $x$ such that $l(u, x) - \sum_i \lambda_i g_i(q_i^\top x) - t > 0$. Therefore, the expectation under this distribution will be positive and the top condition would not hold. This is a contradiction.

Therefore, problem \eqref{eq:innerdual} is equivalent to
\begin{equation} \label{eq:innersip}
\begin{aligned}
    &\min_{\lambda \geq 0, t}\qquad t\\
    &\st \max_{x \in \mathcal{X}}\ l(u, x) - \sum_i \lambda_i g_i(q_i^\top x) -t \leq 0.
\end{aligned}
\end{equation}
Reiterating, under strong duality, \eqref{eq:innersip} is in turn equivalent to \eqref{eq:inner}. We can now include the minimization over $u$ to obtain 
\begin{equation} \label{eq:joint_sip}
\begin{aligned}
    &\min_{u\in \mathcal{U}, \lambda \geq 0, t}\qquad t\\
    &\st\quad l(u, x) - \sum_i \lambda_i g_i(q_i^\top x) -t \leq 0\ \forall\ x \in \mathcal{X},
\end{aligned}
\end{equation}
which is equivalent to \eqref{eq:main}. This is a semi-infinite program \cite{mutapcic2009cutting}. 

In the general case, strong duality between \eqref{eq:main} and \eqref{eq:joint_sip} does not hold. However, by weak duality, \eqref{eq:joint_sip} always constitutes an upper bound of \eqref{eq:main}. Thus, in this paper we solve \eqref{eq:joint_sip} as a (potentially conservative) surrogate for \eqref{eq:main}.

\subsubsection{Semi-Infinite Program for Problem \eqref{eq:prototype}}
We can write problem \eqref{eq:prototype} as 
\begin{equation}\label{eq:prototype_equiv}
\begin{aligned}
    \min_{u\in \mathcal{U}} \max_{\mathbb{P}\in \mathcal{S}}\ &\mathbb{E}_{x \sim \mathbb{P}}\left[ l(u, x) \right]\\
    \mathrm{s.t.}\ &\mathbb{E}_{x \sim \mathbb{P}} \left[ g_i(q_i^\top x) \right] -p_i \leq \epsilon_i,\ \mathrm{for\ all}\ i\\
    &p_i - \mathbb{E}_{x \sim \mathbb{P}} \left[ g_i(q_i^\top x) \right] \leq \epsilon_i,\ \mathrm{for\ all}\ i.
\end{aligned}
\end{equation}
This implies that we can write \eqref{eq:joint_sip} as
\begin{equation}
\begin{aligned}
    &\min_{u\in \mathcal{U}, \lambda^{(1)} \geq 0, \lambda^{(2)} \geq 0, t}\qquad t\\
    &\st \max_{x \in \mathcal{X}}\ l(u, x) - \sum_i (\lambda_i^{(1)}-\lambda_i^{(2)})  g_i(q_i^\top x) +\\
    &\qquad \qquad \sum_i (\lambda_i^{(1)} -\lambda_i^{(2)})p_i +\sum_i (\lambda_i^{(1)} +\lambda_i^{(2)})\epsilon_i-t \leq 0.
\end{aligned}
\end{equation}
By setting $\alpha = \lambda^{(1)} + \lambda^{(2)}$ and $\beta = \lambda^{(1)} - \lambda^{(2)}$, we obtain the equivalent problem
\begin{equation}
\begin{aligned}
    &\min_{u \in \mathcal{U}, \alpha \geq 0, \beta, t}\qquad t\\
    &\st \max_{x \in \mathcal{X}}\ l(u, x) - \sum_i \beta_i  g_i(q_i^\top x) + \sum_i \beta_i p_i +\\
    &\qquad \qquad \sum_i \alpha_i \epsilon_i-t \leq 0\\
    &\qquad \alpha+\beta \geq 0.
\end{aligned}
\end{equation}

\subsection{Optimization Algorithm}
We propose to approximately solve problem \eqref{eq:joint_sip} using an iterative cutting-set method \cite{mutapcic2009cutting}. Our proposed approach is described in \cref{alg:proposed_cutting}. The algorithm iterates between solving a relaxation of problem \eqref{eq:joint_sip}, where $\mathcal{X}$ is replaced by a finite set $\hat{\mathcal{X}}$, and then making the relaxation tighter. The relaxation is tightened by computing a worst-case value for $x$, namely $\hat{x}^*$. The termination criterion implies that upon termination, $(\hat{u}^*, \hat{\lambda}^*, \hat{t}^*)$ is a feasible point of \eqref{eq:joint_sip}.

If $l(u,x)$ is convex in $u$ and $\mathcal{U}$ is a convex set, then we can easily solve problem \eqref{eq:joint_sip_appr} with convex optimization algorithms \cite{boyd2009convex, diamond2016cvxpy}. Further, if $l(u,x)$ is concave in $x$, $\mathcal{X}$ is a convex set, and $g_i$ are convex, the same holds for problem \eqref{eq:inner_cutting}.

If for fixed $x$, $l(u,x)$ can be written as a difference of convex functions in $u$, then problem \eqref{eq:joint_sip_appr} can be solved approximately using disciplined convex-concave programming (DCCP) \cite{shen2016disciplined}. Similarly, we can use DCCP to solve problem \eqref{eq:inner_cutting}, if $l(u,x)$ for fixed $u$ and each function $g_i$ can be written as differences of two convex functions in $x$. If $g_i$ can be written as differences of two convex functions, then $g_i(q_i^\top x)$ can be written as the difference of two convex functions in $x$.  

\begin{algorithm}[t]
\caption{\rev{Cutting-Set} Method}\label{alg:proposed_cutting}
\begin{algorithmic}[1]
\Require discrete set $\hat{\mathcal{X}}=\lbrace x^{(1)}, \dots, x^{(n)}\rbrace \subseteq \mathcal{X}$

\Repeat
\State Solve 
\begin{equation} \label{eq:joint_sip_appr}
\begin{aligned}
    &\min_{u\in \mathcal{U}, \lambda \geq 0, t}\qquad t\\
    &\st\quad l(u, x) - \sum_i \lambda_i g_i(q_i^\top x) -t \leq 0\ \forall x \in \hat{\mathcal{X}},
\end{aligned}
\end{equation}
i.e., problem \eqref{eq:joint_sip} after replacing $\mathcal{X}$ with $\hat{\mathcal{X}}$, and obtain the optimal values for the variables: $\hat{u}^*, \hat{\lambda}^*$, and $ \hat{t}^*$

\State Solve 
\begin{equation}\label{eq:inner_cutting}
    \max_{x \in \mathcal{X}}\ l(\hat{u}^*, x) - \sum_i \hat{\lambda}^*_i g_i(q_i^\top x) - \hat{t}^*
\end{equation}
to obtain $\hat{x}^*$

\State Append $\hat{x}^*$ to $\hat{\mathcal{X}}$
\Until{$l(\hat{u}^*, \hat{x}^*) - \sum_i \hat{\lambda}^*_i g_i(q_i^\top \hat{x}^*) - \hat{t}^* \leq 0$}

\State \Return the optimal control $\hat{u}^*$

\end{algorithmic}
\end{algorithm}

%% file: sections/results.tex
\section{Results}\label{sec:results}
We test our algorithms from \cref{sec:baseline} and \cref{sec:approach} in two scenarios: a portfolio construction problem with application in finance and a trajectory planning problem with application in robotics. 

In order to judge the performance of the algorithms, we choose problems \eqref{eq:protfolio}, \eqref{eq:protfolio_vol}, and \eqref{eq:trajectory}, where we can determine the optimal control by inspection. 
\rev{We note that problem \eqref{eq:protfolio} can be reformulated as a linear program (LP) and solved exactly. However, this does not hold for problem \eqref{eq:protfolio_vol} and problem \eqref{eq:trajectory}. Further, for the last two problems, we cannot use the prior methodology \cite{zhen2023unified} to reformulate them as equivalent finite convex problems, as the necessary conditions are not satisfied. }

\subsection{Portfolio Construction}\label{ref:portf}
We consider the problem of constructing a portfolio with three assets \cite{boyd2024markowitz}. The vector of asset returns is denoted $x \in \mathbb{R}^3$ and we suppose $x\in \left[ -1,1\right]^3$. 
The goal is to determine the optimal portfolio weights, $u\in \mathbb{R}^3$, for a long-only portfolio that maximizes expected return, under uncertainty in the probability distribution of the asset returns $x$. Mathematically,
\begin{equation}\label{eq:protfolio}
\begin{aligned}
    \min_{u \in \mathcal{U}} \ \max_{\mathbb{P}\in \mathcal{S}}\ &\mathbb{E}_{x \sim \mathbb{P}}\left[ -u^\top x \right]\\
    \mathrm{s.t.}\ &|\mathbb{E}_{x \sim \mathbb{P}} \left[ x_1 \right] -0.3| \leq \epsilon_1,\\
    &|\mathbb{E}_{x \sim \mathbb{P}} \left[ x_2 \right] +0.1| \leq \epsilon_2,\\
    &|\mathbb{E}_{x \sim \mathbb{P}} \left[ x_3 \right] - 0.2| \leq \epsilon_3,\\
    &\lVert u \rVert_1 = 1,\ u\geq 0,
\end{aligned}
\end{equation}
where there are constraints on the expected return of each asset and a leverage constraint on the long-only weights. 
\rev{This problem is equivalent to the LP:
\begin{equation}
    \max_{\lVert u \rVert_1 = 1,\ u\geq 0}  (0.3-\epsilon_1)u_1 + (-0.1-\epsilon_2) u_2 + (0.2-\epsilon_3)u_3.
\end{equation}
}

We discuss different cases for $(\epsilon_1, \epsilon_2, \epsilon_3)$ below. For both \cref{alg:baseline} and \cref{alg:proposed_cutting}, the input set $\hat{\mathcal{X}}$ consists of uniformly randomly sampled points in $\left[-1,1 \right]^3$. Problem \eqref{eq:protfolio} results in convex programs within \cref{alg:baseline} and \cref{alg:proposed_cutting}.

We first suppose that $\epsilon_1 = 0.4$, $\epsilon_2 = 0.2$, and $\epsilon_3 = 0.1$.
In this instance of problem \eqref{eq:protfolio}, only asset 3 always has a positive expected return. Under the worst-case distribution for $x$, assets 1 and 2 will have a negative expected return. Therefore, the optimal weight vector is $(0, 0, 1)$. 
This is the weight vector obtained using the cutting-set method, as shown in \Cref{table:portf_constr}. \rev{However, as shown in \Cref{table:portf_constr}, the best-response algorithm gets stuck in a cycle between $(0, 0, 1)$ and $(1, 0, 0)$.} This occurs because a best response (among others) to weight vector $(0, 0, 1)$ is the sample
$x=(0.2, 0.1, 0.1)$, to which the best response is the weight vector $(1, 0, 0)$. 


Second, we consider $\epsilon_1 = 0.2$, $\epsilon_2 = 0.2$, and $\epsilon_3 = 0.1$. In this instance, under the worst-case distribution both assets 1 and 3 have expected return $0.1$, while asset 2 has expected return $-0.1$. Therefore, the optimal weight vectors are of the form $(\tau, 0, 1-\tau)$ for all $\tau \in \left[0, 1\right]$. Both the best-response algorithm and the cutting-set algorithm converge to an optimal weight vector, as shown in \Cref{table:portf_constr}. 

Third, we consider $\epsilon_1=\epsilon_2=\epsilon_3=\ $\SI{e-3}.. Under the worst-case distribution, asset 1 has the best return. Therefore, the optimal weight vector is $(1, 0, 0)$. For small values of $(\epsilon_1, \epsilon_2, \epsilon_3)$, the ambiguity in the distribution of the asset returns decreases. Therefore, we expect it to be more difficult for the best-response algorithm to get trapped in a cycle. We note that in this case both the cutting-set algorithm and the best-response algorithm converge to the optimal weight vector, as shown in \Cref{table:portf_constr}.

Finally, we consider the case $\epsilon_1=\epsilon_2=\epsilon_3=2$. Notice that this implies no actual constraint on the distribution. Any weight vector is identical with respect to the objective of problem \eqref{eq:protfolio}. The different mechanics of the two presented algorithms lead to different, nevertheless optimal, results, as shown in \Cref{table:portf_constr}.

\subsection{Volatility Constraints}
We consider the following extension to problem \eqref{eq:protfolio}:
\begin{equation}\label{eq:protfolio_vol}
\begin{aligned}
    \min_{u \in \mathcal{U}} \ &\max_{\mathbb{P} \in \mathcal{S}}\ \mathbb{E}_{x \sim \mathbb{P}}\left[ -u^\top x \right]\\
    \mathrm{s.t.}\ &|\mathbb{E}_{x \sim \mathbb{P}} \left[ x_1 \right] -0.3| \leq 0.4,\ |\mathbb{E}_{x \sim \mathbb{P}} \left[ x_1^2 \right] -0.34| \leq 0.05,\\
    &|\mathbb{E}_{x \sim \mathbb{P}} \left[ x_2 \right] +0.1| \leq 0.2,\ |\mathbb{E}_{x \sim \mathbb{P}} \left[ x_2^2 \right] -0.26| \leq 0.05,\\
    &|\mathbb{E}_{x \sim \mathbb{P}} \left[ x_3 \right] - 0.2| \leq 0.1,\ |\mathbb{E}_{x \sim \mathbb{P}} \left[ x_3^2 \right] -0.10| \leq 0.05,\\
    &\lVert u \rVert_1 = 1,\ u\geq 0.
\end{aligned}
\end{equation}
Constraining the second central moment of each coordinate of the asset return vector corresponds to constraining the volatility of each asset. This problem leads to a non-convex problem \eqref{eq:sampling_prob2}, for which we use the convex-concave procedure \cite{shen2016disciplined}. Because the objective only depends on the expected asset returns, the second moment constraints are irrelevant. This problem is therefore identical to problem \eqref{eq:protfolio} for $\epsilon_1 = 0.4$, $\epsilon_2 = 0.2$, and $\epsilon_3 = 0.1$. As shown in \Cref{table:portf_constr}, the two algorithms recover the optimal weight vector. Although \eqref{eq:protfolio_vol} is equivalent to \eqref{eq:protfolio}, it is more difficult to solve. In this instance, the best-response algorithm does not get stuck in a cycle, because the extra constraints limit the placement of the samples in \eqref{eq:sampling_prob2}.


\subsection{Trajectory Planning}
We consider the problem of controlling a linear system of the form 
\begin{equation}
    x_{t+1} = \underbrace{\begin{bmatrix}
        0.4 & 1.5\\
        0 &0.9
    \end{bmatrix}}_{A} x_t + \underbrace{\begin{bmatrix}
        0\\
        1 
    \end{bmatrix}}_{B} u_t\ \mathrm{for\ all}\ t \geq 0,
\end{equation}
where $x_t \in \mathbb{R}^2$ is the state and $u_t \in \mathbb{R}$ the control at time $t$. Notice that for any $t >0$, using recursion, we can write
\begin{equation}
    x_t = A^t x_0 + \sum_{\tau =0}^{t-1}A^{t-1-\tau}B u_\tau.
\end{equation}
We assume that the initial state $x_0$ is unknown but lies in $\mathcal{X} = \left[ -0.3,0.3\right]^2$.
We consider the problem of reaching a target position, $z_\mathrm{goal}=(0, 0)$, at time $T$, under uncertain initial state $x_0$:
\begin{equation}\label{eq:trajectory}
\begin{aligned}
    \min_{u_0, \dots, u_{T-1} \in \mathbb{R}} \ \max_{\mathbb{P}\in \mathcal{S}}\ &\mathbb{E}_{x_0 \sim \mathbb{P}} \lVert z_T - z_\mathrm{goal}\rVert_2 \\
    \mathrm{s.t.}\ &|\mathbb{E}_{x_0 \sim \mathbb{P}} \left[ x_1 \right] -0.1| \leq 0.1,\\
    &|\mathbb{E}_{x_0 \sim \mathbb{P}} \left[ x_2 \right] + 0.1| \leq 0.2,\\
    & u_t \in \left[-0.1, 0.1\right]\ \mathrm{for\ all}\ t=0, \dots, T-1.
\end{aligned}
\end{equation}

By computing the eigenvalues of $A$, we conclude that the linear dynamical system is asymptotically stable, i.e., the state converges to the origin as time progresses, without any input. Therefore, assuming $T \rightarrow \infty$, an optimal open-loop control sequence for \eqref{eq:trajectory} is the zero input $\mathbf{0}$. 

\begin{figure}[h]
    \centering
    \begin{subfigure}[b]{0.50\textwidth}
        \centering
        \includegraphics[width=\textwidth]{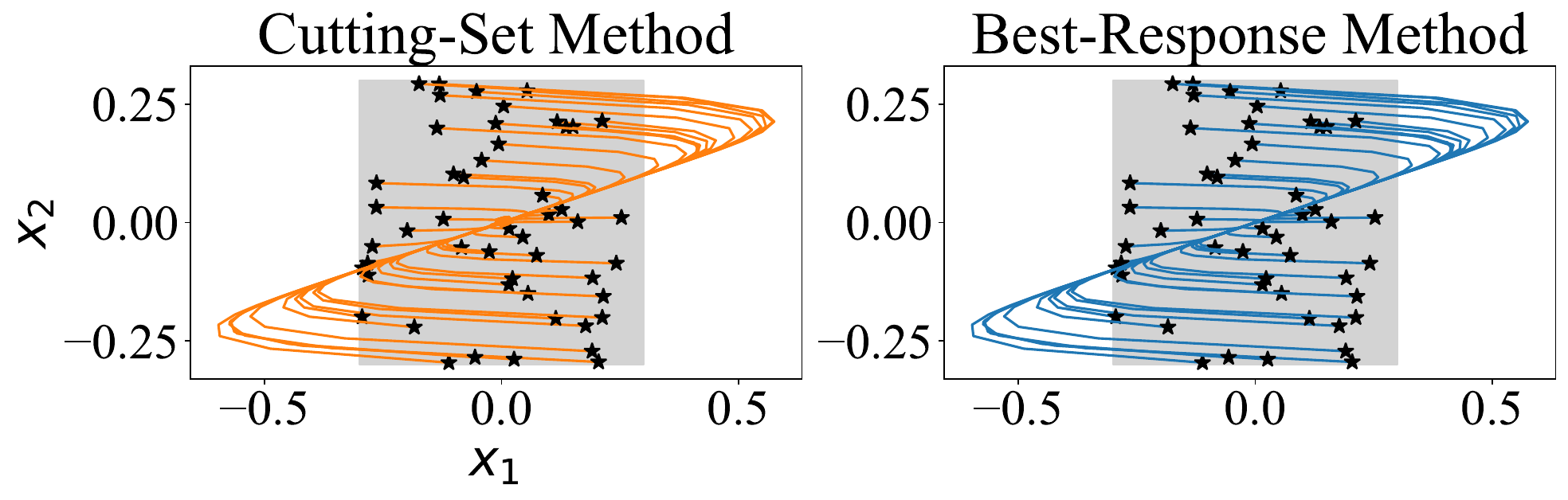}
        \caption{$T=50, z_\mathrm{goal}=(0, 0)$.}
        \label{fig:t50}
    \end{subfigure}
    \hfill
    \begin{subfigure}[b]{0.50\textwidth}
        \centering
        \includegraphics[width=\textwidth]{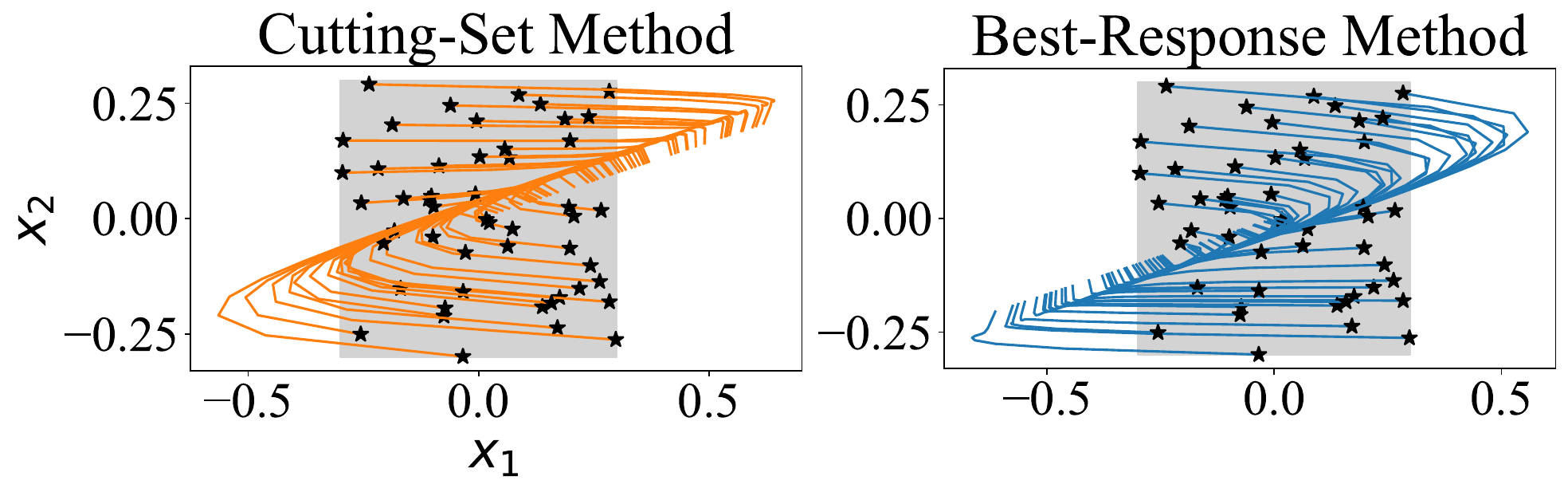}
        \caption{$T=10, z_\mathrm{goal}=(0, 0)$.}
        \label{fig:t10}
    \end{subfigure}

        \begin{subfigure}[b]{0.3\textwidth}
        \centering
        \includegraphics[width=\textwidth]{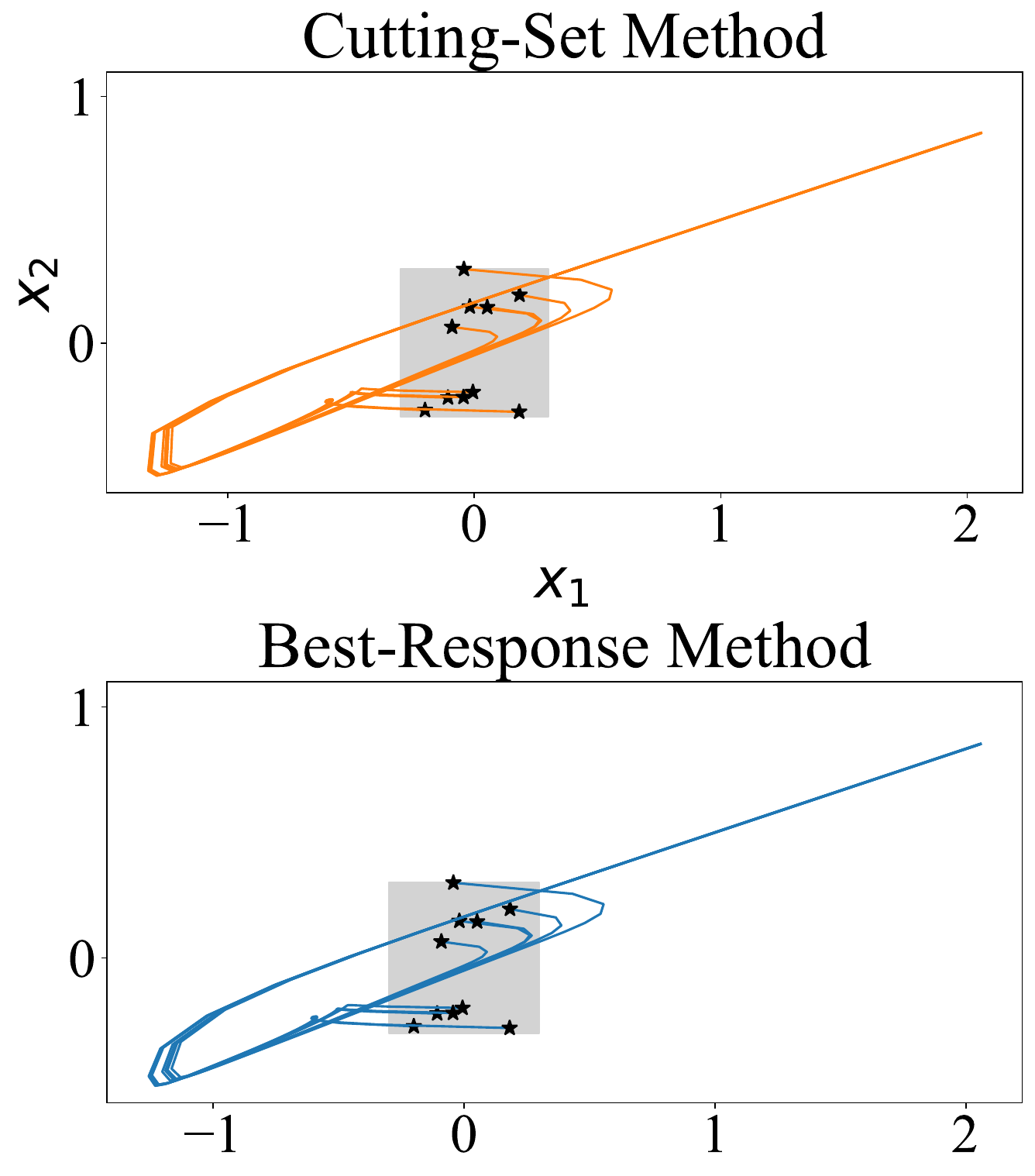}
        \caption{$T=50, z_\mathrm{goal}=(2, 1)$.}
        \label{fig:t50_21}
    \end{subfigure}
    \caption{Trajectories for randomly sampled initial state $x_0 \in \left[-0.3, 0.3\right]^2$ and open-loop control as determined by \Cref{alg:baseline} and \Cref{alg:proposed_cutting} for problem \eqref{eq:trajectory}. The asterisk denotes the initial state for each trajectory. }
    \label{fig:traj}
\end{figure}

This is in fact the control input obtained by the two methods for the case of large $T$ ($T=50$). In \Cref{fig:t50}, we observe that independently of the initial state, which is randomly sampled in $\left[-0.3, 0.3 \right]^2$, the state trajectory converges close to the origin. In the case of a smaller $T$ ($T=10$), the obtained control input is no longer zero. The shorter horizon requires a nonzero control to force the trajectories close to the origin. In this case, as shown in \Cref{fig:t10}, it is more difficult to drive all initial states to the origin, with some trajectories ending farther away.

\rev{We further consider the case of $T=50$ and $z_\mathrm{goal} = (2,1)$. The optimal control is no longer $\mathbf{0}$, but as shown in \Cref{fig:t50_21}, both algorithms drive the state close to $(2,1)$ for randomly sampled initial states in $\left[-0.3, 0.3 \right]^2$. We include a smaller number of sampled initial states to improve the clarity of the trajectories. }

Finally, note that in this set-up, problems \eqref{eq:sampling_prob2} and \eqref{eq:inner_cutting} are not convex. We approximately solve them using DCCP.
\rev{In the trajectory planning set-up, which is the most complicated from the scenarios considered, \revf{for the two algorithms}, we \revf{report} the \revf{average} run-time \revf{for a single iteration of their main loop}. The implementation is in Python, using the packages CVXPY \cite{diamond2016cvxpy} and DCCP \cite{shen2016disciplined}. We run the algorithms on an Apple M1 Pro chip and 32GM of memory. \revf{The cutting-set algorithm took \SI{0.29}{\second} on average for a single iteration of the main loop, while the best-response algorithm required \SI{0.51}{\second}}. The cutting-set method is significantly faster, because the size of the involved problems is smaller. Note that the best-response algorithm needs to update the $n$ points in $\mathcal{X}$, by solving \eqref{eq:sampling_prob2}, in every iteration.
\revf{In more detail, a convex problem with $k$ variables and $m$ constraints can be solved in roughly $\mathcal{O}\left(\sqrt{m}(k+m)^3\right)$ flops using an interior-point method \cite{boyd2009convex}. Assuming both algorithms involve convex problems, we conclude that the best-response algorithm has a higher complexity. It is therefore reasonable that, for the non-convex case where we apply DCCP, the best-response method also has a higher complexity.}
}


\begin{table*}[h!]
\centering
\caption{Weight vector solution $u^\star$ for the portfolio construction problems \eqref{eq:protfolio} and \eqref{eq:protfolio_vol} for different selections of $(\epsilon_1, \epsilon_2, \epsilon_3)$. Each row corresponds to an algorithm. A double-sided arrow $\leftrightarrow$ for the weight vector by the best-response algorithm indicates a cycle.}
\label{table:portf_constr}
\begin{tabular}{
    >{\bfseries}c c c c c c
}
\toprule
\textbf{} & \multicolumn{4}{c}{\textbf{Problem \eqref{eq:protfolio}}} & \textbf{Problem \eqref{eq:protfolio_vol}} \\
\cmidrule(lr){2-5} \cmidrule(l){6-6}
\textbf{$(\epsilon_1, \epsilon_2, \epsilon_3)$} & $(0.4, 0.2, 0.1)$ & $(0.2, 0.2, 0.1)$ & $(10^{-3}, 10^{-3}, 10^{-3})$ & $(2.0,2.0,2.0)$ & $(0.4, 0.2, 0.1)$ \\
\midrule
\textbf{Best Response} & $(0.0,0.0,1.0) \leftrightarrow (1.0,0.0,0.0)$ & $(0.67, 0.0, 0.33)$ & $(1.0, 0.0, 0.0)$ & $(0.7, 0.0, 0.3)$ & $(0.0, 0.0, 1.0)$ \\
\textbf{Cutting-Set} & $(0.0,0.0,1.0)$ & $(0.48, 0.0, 0.52)$ & $(1.0, 0.0, 0.0)$ & $(0.37, 0.36, 0.27)$ & $(0.0, 0.0, 1.0)$\\
\bottomrule
\end{tabular}
\end{table*}

%% file: sections/conclusion.tex
\section{Conclusion}\label{sec:conclusion}
We present methods for a class of problems in distributionally robust control where the ambiguity set is defined using 1-dimensional linear projections of the uncertain parameters. This problem class is general and includes problems with interpretable ambiguity sets. Our first method is based on the idea of best-responses from game theory. The second method is based on duality theory and cutting-set methods. We consider simple experimental set-ups and show that the method based on duality theory converges to the optimal control. The method based on best-response dynamics obtained similar convergence in most cases, but it can get trapped in a cycle. Nevertheless, the method based on duality theory also has limitations: it is only exact if strong duality holds and the proposed cutting-set method converges to the solution of problem \eqref{eq:joint_sip}. \rev{This method is inferior to the method by Zhen et al. \cite{zhen2023unified}, if the necessary conditions that allow for an equivalent finite convex formulation hold, but it can be used to obtain an approximate solution in the general case, where the method by Zhen et al. is not applicable}. Future work will deal with the theoretical analysis of both methods \revf{that could rely on game theory and fixed point theorems for the best-response method and convex analysis and duality theory for the cutting-set algorithm}. Further, an investigation of the cases for which the best-response method gets trapped in a cycle is interesting.
